\documentclass[a4paper,draft]{amsart}
\makeatletter
\@addtoreset{equation}{section}
\makeatother

\newtheorem{proposition}{Proposition}

\theoremstyle{definition}

\theoremstyle{remark}

\DeclareMathOperator{\Ad}{Ad}

\newcommand{\cla}{c_b}

\newcommand{\DEHN}{\mathsf D}

\newcommand{\FUNCTOR}{\mathsf F}
\newcommand{\GEN}{\mathsf f}
\newcommand{\GENI}{\mathsf g}

\newcommand{\IMUN}{\mathsf i}
\newcommand{\INCL}{\kappa}
\newcommand{\INTEGERS}{\mathbb Z}
\newcommand{\la}{b}
\newcommand{\LC}{\mathsf U_{lc}}

\newcommand{\MOM}{\mathsf p}

\newcommand{\OBALG}{\mathcal A_{N}}
\newcommand{\PERMUTE}{\mathsf P}

\newcommand{\POS}{\mathsf q}
\newcommand{\PTOLEMY}{\mathsf T}
\newcommand{\QDILOG}{\mathop{\varphi_b}}
\newcommand{\QDILOGI}{\mathop{\bar{\varphi}_b}}

\newcommand{\REALS}{\mathbb R}
\newcommand{\ROTATE}{\mathsf A}

\newcommand{\SHFL}{\mathsf G}

\newcommand{\SIM}{\mathsf W}

\newcommand{\Spec}{\mathop{\mathrm{Spec}}}

\newcommand{\TLC}{\tilde {\mathsf U}_{lc}}

\title[Strongly coupled quantum discrete Liouville theory]{Strongly
coupled quantum discrete Liouville theory. II:
Geometric interpretation of the evolution operator}
\author{L.D. Faddeev}
\address{Steklov Mathematical Institute at St. Petersburg,
Fontanka 27, St. Petersburg 191011, Russia}
\email{faddeev@pdmi.ras.ru, kashaev@pdmi.ras.ru}
 \author{R.M. Kashaev}

\date{January 2002}
\keywords{discrete Liouville equation, quantum integrable systems,
  quantum Teichm\"uller theory.\\
\indent The work is supported in part by grants CRDF RM1-2244, INTAS
  99-01705, and RFBR 
  99-01-00101}
\begin{document}
\begin{abstract}
  It is shown that the $N$-th power of the light-cone
  evolution operator of $2N$-periodic quantum discrete Liouville model
can be identified with the 
Dehn twist operator in quantum Teichm\"uller theory. 
\end{abstract}
\maketitle

\section*{Introduction}

Integrable lattice regularization
of quantum Liouville theory has been developed in
papers \cite{fadtak,faddeev91,fadVol97}. According to recent development
in \cite{fkv}, the model is expected to describe quantum
Liouville equation with Virasoro central charge $c_L>1$, including
the ``strongly coupled regime'' $1<c_L<25$.
 
This paper can be considered as a second part of our previous work \cite{fkv} 
dedicated to discrete Liouville model. Here we show that the evolution
operator of the 
model can be interpreted in pure geometrical terms within
quantum Teichm\"uller theory
\cite{fock2,fock1,kash1,kash2,kash3}. 
Namely, we identify  the $N$-th power of the
light-cone evolution operator of quantum discrete Liouville model of
spatial length $2N$ (which is the number of sites in a chain) with the
Dehn twist operator in quantum Teichm\"uller theory. 
 
The paper is organized as follows.  The quantum discrete Liouville
system is briefly described in Section~\ref{sec:qdls}.  The
relation to quantum Teichm\"uller theory is explained in
Section~\ref{sec:qdl-qtt}. 
\subsection*{Acknowledgements}
The authors would like to thank A. Fring, M. Karowski,
   R. Schrader, S. Sergeev, Yu. Suris, J. Teschner
   for useful discussions.

\section{Quantum discrete Liouville system}
\label{sec:qdls}

\subsection{Algebra of observables}
Following \cite{fkv}, algebra of observables $\OBALG$, $N> 1$, is 
generated by selfadjoint elements
$\GEN_j$, $j\in\INTEGERS$, with periodicity condition
\(
\GEN_{j+2N}=\GEN_j
\)
and commutation relations
\begin{equation}\label{eq:com-rel}
[\GEN_m,\GEN_{n}]=\left\{
\begin{array}{cl}
(-1)^m(2\pi\IMUN)^{-1},&\mathrm{if}\ n=m\pm1\pmod{2N}\\
0,&\mathrm{otherwise}
\end{array}\right.
\end{equation}
\subsection{Equations of motion}
The field variables 
\[
\chi_{j,t}\equiv \LC^t e^{2\pi\la\GEN_{j+t}}\LC^{-t},\quad j+t=1\pmod2
\]
are defined so that
\[
\LC\chi_{j,t}\LC^{-1}=\chi_{j-1,t+1}
\]
Here
the ``light-cone'' evolution operator $\LC$ is defined explicitly
\begin{equation}
\label{eq:lc}
\LC=\SHFL\prod_{j=1}^N\QDILOG(\GEN_{2j})
=\prod_{j=1}^N\QDILOG(\GEN_{2j-1})\SHFL 
\end{equation}
where
\begin{multline*}
\QDILOG(z)\equiv\exp\left(\frac{1}{4}
\int_{\IMUN 0-\infty}^{\IMUN 0+\infty}
\frac{e^{-\IMUN 2 zw}\, dw}{\sinh(w\la)
\sinh(w/\la) w}\right)\\
=
(e^{2\pi (z+\cla)\la};q^2)_\infty/
(e^{2\pi (z-\cla)\la^{-1}};\bar q^2)_\infty,
\end{multline*}
\[
q\equiv e^{\IMUN\pi\la^2},\quad
\bar q\equiv e^{-\IMUN\pi\la^{-2}},\quad \cla\equiv\IMUN(\la+\la^{-1})/2
\]
while operator $\SHFL$ is defined through the equations
\begin{equation}\label{eq:shfl}
\SHFL\GEN_j=(-1)^{j}\GEN_{j-1}\SHFL
\end{equation}
The field variables solve the 
quantum discrete Liouville equation
\footnote{Using invariance of $\LC$ with respect to 
symmetry $\la\leftrightarrow\la^{-1}$, 
one can also define the dual fields satisfying the dual equation, see 
\cite{fkv}.}
\[
\chi_{j,t+1}\chi_{j,t-1}
   =(1+q\chi_{j+1,t})(1+q\chi_{j-1,t})
\]
with spatial periodic boundary condition
\[
\chi_{j+2N,t}=\chi_{j,t}
\]
and the initial data given by exponentiated generators
\[
\chi_{2j+1,0}=e^{2\pi\la\GEN_{2j+1}},\quad 
\chi_{2j,-1}=e^{2\pi\la\GEN_{2j}}
\]

\section{Interpretation within quantum Teichm\"uller theory}
\label{sec:qdl-qtt}
In this section we interpret the evolution operator $\LC$ in geometrical 
terms by
using the formalism of
decorated ideal triangulations and 
their transformations within quantum 
Teichm\"uller theory
described in \cite{kash3}.

\subsection{Geometric realization}
\begin{figure}[htb]
\centering
\begin{picture}(200,40)
\put(0,0){\begin{picture}(120,40)
\multiput(0,0)(0,40){2}{\line(1,0){120}}
\multiput(0,0)(60,0){3}{\line(0,1){40}}
\multiput(0,40)(60,0){2}{\line(3,-2){60}}
\multiput(0,0)(60,0){3}{\circle*{3}}
\multiput(0,40)(60,0){3}{\circle*{3}}
\scriptsize
\multiput(1,32)(60,0){2}{$*$}
\multiput(55,3)(60,0){2}{$*$}
\put(15,10){1}
\put(75,10){3}
\put(40,25){2}
\put(100,25){4}
\end{picture}}

\put(140,0){\begin{picture}(60,40)
\multiput(0,0)(0,40){2}{\line(1,0){60}}
\multiput(0,0)(60,0){2}{\line(0,1){40}}
\multiput(0,40)(60,0){1}{\line(3,-2){60}}
\multiput(0,0)(60,0){2}{\circle*{3}}
\multiput(0,40)(60,0){2}{\circle*{3}}
\scriptsize
\multiput(1,32)(60,0){1}{$*$}
\multiput(55,3)(60,0){1}{$*$}
\put(4,10){$2N-1$}
\put(40,25){$2N$}
\end{picture}}

\multiput(125,0)(5,0){3}{\circle*{1}}
\multiput(125,40)(5,0){3}{\circle*{1}}
\end{picture}
\caption{Decorated ideal triangulation $\tau_N$
of an annulus with $N$ marked 
points on each boundary component. 
The leftmost and the rightmost vertical 
edges are identified.}\label{fig:1}
\end{figure}
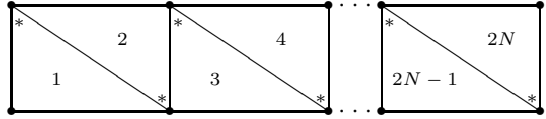

We consider an annulus with $N$ marked points on each of its 
boundary components ($2N$ points in total) 
and choose decorated ideal triangulation $\tau_N$
of it shown in Fig.~\ref{fig:1}. 
Equivalently, we can think of $\tau_N$ as an infinite triangulated
strip where triangles are numerated by integers in accordance 
with Fig.~\ref{fig:1} with periodicity condition
\(
\bar\tau_N(n+2N)=\bar\tau_N(n),\ \forall n\in\INTEGERS
\).
In this way we come to identification of the integers
from $1$ to $2N$, numbering triangles in $\tau_N$,
with  elements of the ring of residues 
\(
\INTEGERS_{2N}\equiv\INTEGERS/2N\INTEGERS
\). 
Such identification will be assumed in algebraic expressions, when necessary.

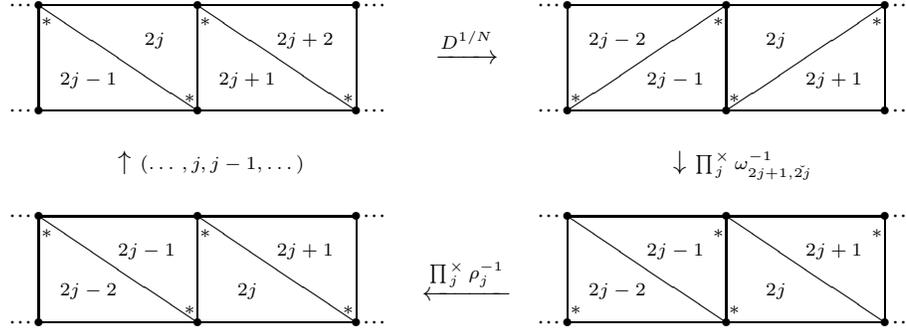
\begin{figure}[htb]
\centering
\begin{picture}(320,120)
\put(0,80){\begin{picture}(120,40)
\multiput(0,0)(0,40){2}{\line(1,0){120}}
\multiput(0,0)(60,0){3}{\line(0,1){40}}
\multiput(0,40)(60,0){2}{\line(3,-2){60}}
\multiput(0,0)(60,0){3}{\circle*{3}}
\multiput(0,40)(60,0){3}{\circle*{3}}
\scriptsize
\multiput(1,32)(60,0){2}{$*$}
\multiput(55,3)(60,0){2}{$*$}
\put(8,10){$2j-1$}
\put(68,10){$2j+1$}
\put(40,25){$2j$}
\put(90,25){$2j+2$}
\multiput(-4,0)(-3,0){3}{\circle*{1}}
\multiput(-4,40)(-3,0){3}{\circle*{1}}
\multiput(124,0)(3,0){3}{\circle*{1}}
\multiput(124,40)(3,0){3}{\circle*{1}}
\end{picture}}
\put(200,80){\begin{picture}(120,40)
\multiput(0,0)(0,40){2}{\line(1,0){120}}
\multiput(0,0)(60,0){3}{\line(0,1){40}}
\multiput(0,0)(60,0){2}{\line(3,2){60}}
\multiput(0,0)(60,0){3}{\circle*{3}}
\multiput(0,40)(60,0){3}{\circle*{3}}
\scriptsize
\multiput(1,3)(60,0){2}{$*$}
\multiput(55,32)(60,0){2}{$*$}
\put(8,25){$2j-2$}
\put(75,25){$2j$}
\put(30,10){$2j-1$}
\put(90,10){$2j+1$}
\multiput(-4,0)(-3,0){3}{\circle*{1}}
\multiput(-4,40)(-3,0){3}{\circle*{1}}
\multiput(124,0)(3,0){3}{\circle*{1}}
\multiput(124,40)(3,0){3}{\circle*{1}}
\end{picture}}
\put(0,0){\begin{picture}(120,40)
\multiput(0,0)(0,40){2}{\line(1,0){120}}
\multiput(0,0)(60,0){3}{\line(0,1){40}}
\multiput(0,40)(60,0){2}{\line(3,-2){60}}
\multiput(0,0)(60,0){3}{\circle*{3}}
\multiput(0,40)(60,0){3}{\circle*{3}}
\scriptsize
\multiput(1,32)(60,0){2}{$*$}
\multiput(55,3)(60,0){2}{$*$}
\put(8,10){$2j-2$}
\put(75,10){$2j$}
\put(30,25){$2j-1$}
\put(90,25){$2j+1$}
\multiput(-4,0)(-3,0){3}{\circle*{1}}
\multiput(-4,40)(-3,0){3}{\circle*{1}}
\multiput(124,0)(3,0){3}{\circle*{1}}
\multiput(124,40)(3,0){3}{\circle*{1}}
\end{picture}}
\put(200,0){\begin{picture}(120,40)
\multiput(0,0)(0,40){2}{\line(1,0){120}}
\multiput(0,0)(60,0){3}{\line(0,1){40}}
\multiput(0,40)(60,0){2}{\line(3,-2){60}}
\multiput(0,0)(60,0){3}{\circle*{3}}
\multiput(0,40)(60,0){3}{\circle*{3}}
\scriptsize
\multiput(1,3)(60,0){2}{$*$}
\multiput(55,32)(60,0){2}{$*$}
\put(8,10){$2j-2$}
\put(75,10){$2j$}
\put(30,25){$2j-1$}
\put(90,25){$2j+1$}
\multiput(-4,0)(-3,0){3}{\circle*{1}}
\multiput(-4,40)(-3,0){3}{\circle*{1}}
\multiput(124,0)(3,0){3}{\circle*{1}}
\multiput(124,40)(3,0){3}{\circle*{1}}
\end{picture}}
\put(150,90){$\stackrel{D^{1/N}}{\overrightarrow{\phantom{D^{1/N}}}}$}
\put(145,0){$\stackrel{\prod^{\times}_{j}
\rho^{-1}_j}{\overleftarrow{\phantom{\prod^{\times}_{j}
\rho^{-1}_j}}}$}
\put(240,58){$\downarrow$ {\scriptsize $\prod^{\times}_{j} 
\omega_{2j+1,\check{2j}}^{-1}$}}
\put(30,58){$\uparrow$ {\scriptsize $(\ldots,j,j-1,\ldots)$}}
\end{picture}
\caption{Continuous transformation $D^{1/N}$ of triangulated
annulus $\tau_N$ 
is a cyclic shift of the
top boundary wrt the bottom boundary to the right by one spacing.}\label{fig:2}
\end{figure}

Denote $D^{1/N}$
the isotopy class of a homeomorphism of the annulus
which rotates the top boundary component wrt the bottom thru angle
$2\pi /N$ so that the marked points of the top boundary are cyclically 
shifted by one period. The reason for using fractional power notation comes
from the fact that 
\[
\underbrace{D^{1/N}\circ\dots\circ  D^{1/N}}_{N\ \mathrm{times}}=D
\]
is nothing else but the Dehn twist.
From Fig.~\ref{fig:2} 
it follows that the following composition of geometric transformations is
identity,
\[
(\ldots,j,j-1,\ldots)\circ \prod\nolimits^{\times}_{k}
\rho^{-1}_k\circ\prod\nolimits^{\times}_{l} 
\omega_{2l+1,\check{2l}}^{-1}\circ D^{1/N}=id
\]
where elementary geometric transformations $\rho_i$ and $\omega_{ij}$ have
the form
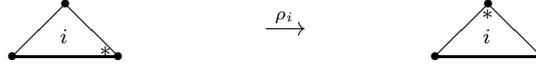
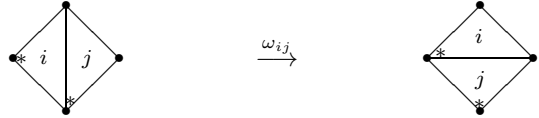
\begin{center}
\begin{picture}(200,20)
\put(0,0){\begin{picture}(40,20)
\put(0,0){\line(1,0){40}}
\put(0,0){\line(1,1){20}}
\put(20,20){\line(1,-1){20}}
\put(0,0){\circle*{3}}
\put(20,20){\circle*{3}}
\put(40,0){\circle*{3}}
\footnotesize
\put(33,0){$*$}
\put(18,5){$i$}
\end{picture}}
\put(160,0){\begin{picture}(40,20)
\put(0,0){\line(1,0){40}}
\put(0,0){\line(1,1){20}}
\put(20,20){\line(1,-1){20}}
\put(0,0){\circle*{3}}
\put(20,20){\circle*{3}}
\put(40,0){\circle*{3}}
\footnotesize
\put(17.5,14){$*$}
\put(18,5){$i$}
\end{picture}}
\put(95,8){$\stackrel{\rho_i}{\longrightarrow}$}
\end{picture}
\end{center}
and
\begin{center}
\begin{picture}(200,40)
\put(0,0){
\begin{picture}(40,40)
\put(20,0){\line(-1,1){20}}
\put(40,20){\line(-1,-1){20}}
\put(0,20){\line(1,1){20}}
\put(40,20){\line(-1,1){20}}
\put(20,0){\line(0,1){40}}
\put(20,0){\circle*{3}}
\put(0,20){\circle*{3}}
\put(20,40){\circle*{3}}
\put(40,20){\circle*{3}}
\footnotesize
\put(10,18){$i$}\put(26,18){$j$}
\put(1,18){$*$}
\put(19.5,2){$*$}
\end{picture}}
\put(160,0){\begin{picture}(40,40)
\put(20,0){\line(-1,1){20}}
\put(40,20){\line(-1,-1){20}}
\put(0,20){\line(1,1){20}}
\put(40,20){\line(-1,1){20}}
\put(0,20){\line(1,0){40}}
\put(20,0){\circle*{3}}
\put(0,20){\circle*{3}}
\put(20,40){\circle*{3}}
\put(40,20){\circle*{3}}
\footnotesize
\put(18,26){$i$}\put(18,10){$j$}
\put(3,20){$*$}
\put(17.5,1){$*$}
\end{picture}}
\put(95,17){$\stackrel{\omega_{ij}}{\longrightarrow}$}
\end{picture}
\end{center}
with
\[
\omega_{k,\check l}\equiv \rho_l\circ\omega_{k,l}\circ\rho_l^{-1}
\]
while $(\ldots,j,j-1,\ldots)$ denotes the index shift transformation 
$j\mapsto j-1$.  
Equivalently we can write
\[
D^{1/N}=\prod\nolimits^{\times}_{l} 
\omega_{2l+1,\check{2l}}\circ\prod\nolimits^{\times}_{k}
\rho_k\circ(\ldots,j,j+1,\ldots)
\]
so, quantum realization of $D^{1/N}$ in $L^2(\REALS^{2N})$ has the form 
\begin{equation}\label{eq:d-1/N}
\FUNCTOR\left(\tau_N,D^{1/N}(\tau_N)\right)\simeq \DEHN^{1/N}
\equiv
\zeta^{-N-6/N}\PERMUTE_{(\ldots j,{j+1}\ldots)}
\prod_{k=1}^{2N}\ROTATE_k
\prod_{l=1}^N\PTOLEMY_{2l+1,\check{2l}}
\end{equation}
with the normalization factor chosen in accordance with the convention
for Dehn twists used  in \cite{kash2}.
Here $\PERMUTE_{(\ldots j,{j+1}\ldots)}$ is the natural realization of the 
cyclic permutation,
\[
\zeta=e^{-\IMUN\pi(\la+\la^{-1})^2/12}
\]
\[
 \ROTATE_k\equiv e^{-\IMUN\pi/3}e^{\IMUN
    3\pi\POS^2_k}e^{\IMUN\pi(\MOM_k+\POS_k)^2}
\]
\[
\PTOLEMY_{k,\check
  l}=e^{-\IMUN2\pi\MOM_k\MOM_l}\QDILOGI(\POS_k+\POS_l),\quad
  \QDILOGI(z)\equiv (\QDILOG(z))^{-1}
\]
where selfadjoint operators $\MOM_j,\POS_j$ satisfy Heisenberg 
commutation relations
\[
[\MOM_j,\MOM_k]=[\POS_j,\POS_k]=0,\quad [\MOM_j,\POS_k]=\delta_{j,k}
(2\pi\IMUN)^{-1}
\]
We can rewrite eqn~\eqref{eq:d-1/N} in the form
\begin{equation}\label{eq:d-1/N-1}
\DEHN^{-1/N}=\zeta^{N+6/N}\prod_{m=1}^N\QDILOG(\POS_{2m}+\POS_{2m+1})
e^{\IMUN2\pi\sum_{j=1}^N\MOM_{2j}\MOM_{2j+1}}
\prod_{k=1}^{2N}\ROTATE_k^{-1}
\PERMUTE_{(\ldots,l,{l-1},\ldots)}
\end{equation}
\begin{proposition}
Operators
\begin{equation}\label{eq:rfj}
\INCL(\GEN_j)=\left\{
\begin{array}{cl}
\MOM_j+\MOM_{j-1},&\mathrm{if}\ j=0\pmod2\\
\POS_j+\POS_{j-1},&\mathrm{otherwise}
\end{array}\right.
\end{equation}
\begin{equation}\label{eq:revol}
\INCL(\SHFL)=\zeta^{N+6/N}e^{\IMUN2\pi\sum_{j=1}^N\MOM_{2j}\MOM_{2j+1}}
\prod_{k=1}^{2N}\ROTATE_k^{-1}
\PERMUTE_{(\ldots,l,{l-1},\ldots)}
\end{equation}
define faithful (reducible) realization of the 
observable algebra $\OBALG$  in $L^2(\REALS^{2N})$.
\end{proposition}
\begin{proof}
Let us check that these definitions are consistent with
relations~\eqref{eq:com-rel}, \eqref{eq:shfl}. First, evidently,
\[
[\INCL(\GEN_{2j}),\INCL(\GEN_{2k})]=[\INCL(\GEN_{2j+1}),
\INCL(\GEN_{2k+1})]=0
\]
while
\[
[\INCL(\GEN_{2j}),\INCL(\GEN_{2k+1})]=[\MOM_{2j}+\MOM_{2j-1},
\POS_{2k+1}+\POS_{2k}]=(2\pi\IMUN)^{-1}(\delta_{j,k}+\delta_{j,k+1}) 
\]
thus reproducing relations~\eqref{eq:com-rel}. Next, 
\begin{multline*}
\Ad(\INCL(\SHFL))\INCL(\GEN_{2j})=\Ad(\INCL(\SHFL))
(\MOM_{2j}+\MOM_{2j-1})=\Ad(e^{\IMUN
  2\pi\sum_{k=1}^N\MOM_{2k}\MOM_{2k+1}})(\MOM_{\check{2j-1}}
+\MOM_{\check{2j-2}})\\
=\Ad(e^{\IMUN
  2\pi\sum_{k=1}^N\MOM_{2k}\MOM_{2k+1}})(\POS_{2j-1}-\MOM_{2j-1}
+\POS_{2j-2}-\MOM_{2j-2})=\POS_{2j-1}+\POS_{2j-2}=\INCL(\GEN_{2j-1})
\end{multline*}
and similarly
\begin{multline*}
\Ad(\INCL(\SHFL))\INCL(\GEN_{2j+1})=\Ad(\INCL(\SHFL))
(\POS_{2j+1}+\POS_{2j})\\=\Ad(e^{\IMUN
  2\pi\sum_{k=1}^N\MOM_{2k}\MOM_{2k+1}})(\POS_{\check{2j}}
+\POS_{\check{2j-1}})
=-\MOM_{2j}-\MOM_{2j-1}=-\INCL(\GEN_{2j})
\end{multline*}
in agreement with eqn~\eqref{eq:shfl}.
\end{proof}
Now, comparing eqns~\eqref{eq:lc}, \eqref{eq:d-1/N-1}, we come to our
main result
\begin{equation}\label{eq:u-d}
\INCL(\LC)=\DEHN^{-1/N}
\end{equation}
\subsection{A similarity transformation}
Here, we give (without proof) the result of similarity transformation which 
simplifies the $N$-th power of the evolution operator.

Define
\[
\SIM\equiv\prod_{N\ge j>1}\left(\QDILOGI(\GEN_{2j})\QDILOG(\GEN_{2j-1})
\QDILOG(\GENI_{2j,2N+1})\right),\quad 
\QDILOGI(x)\equiv (\QDILOG(x))^{-1}
\]
where
\(
\GENI_{j,k}\equiv\sum_{l=j+1}^k\GEN_{j}
\), and the product of noncommuting operators is in decreasing order 
from left to right.
\begin{proposition}
One has the following explicit expression for the 
transformed evolution operator
\begin{multline*}
\TLC\equiv \Ad(\SIM^{-1})\LC=(\zeta e^{\IMUN\pi/6})^{-N}
\prod_{N>k>1}\QDILOGI(\GENI_{2k-1,2N-1})\\
\times
\QDILOGI(\GENI_{2,2N-1})\QDILOG(\GENI_{2N-1,2N+1})
\QDILOGI(\GEN_{2})e^{\IMUN\pi\sum_{l=1}^N\GEN_{2l}^2}\SHFL
\end{multline*}
where the product is again in decreasing order 
from left to right, while the $N$-th power has the form
\[
\TLC^{N}=(\zeta e^{\IMUN\pi/6})^{1-N^2}
\QDILOG(\GENI_{2,2N+1})e^{-\IMUN\pi\GEN_2^2}
\left( e^{\IMUN\pi\sum_{l=1}^N\GEN_{2l}^2}\SHFL\right)^{N}
\] 
\end{proposition}
\section*{Conclusion}

The main result of this paper is formula \eqref{eq:u-d}
which, on one hand side, identifies ``zero-modes'' of $2N$-periodic
quantum discrete 
Liouville equation to be given by $N$-th power of the light-cone
evolution operator $\LC$, and equates these
zero-modes to the (inverse 
of) Dehn twist operator in quantum Teichm\"uller theory, on the other.
Immediate consequence of this result is that now, based on the
known spectrum of operator $\DEHN$, we know the
spectrum of the model. Indeed, the 
complete spectrum of $\DEHN$ is given by the formula \cite{kash3}:
\[
\Spec(\DEHN)=\{e^{\IMUN2\pi\Delta_s}|s\in\REALS_{>0}\}
\]
where
\[
\Delta_s=\frac{c_L-1}{24}+s^2,\quad c_L=1+6(\la+\la^{-1})^2>1
\]
are conformal weights and the Virasoro central charge in (continuous) quantum
Liouville theory, see \cite{tesch} for a recent review. 
This is consistent with interpretation of the Dehn
twist spectrum as Liouville conformal weights through the formula
$\Spec(\DEHN)=\Spec( e^{\IMUN2\pi L_0})$, where $L_0$ is the Virasoro
generator in continuous quantum Liouville theory with the known
spectrum
\[
\Spec(L_0)=\{\Delta_s+m|s\in\REALS_{>0},\ m\in\INTEGERS_{\ge 0}\}
\]
Our result implies the following spectrum of $\LC$:
\[
\Spec(\LC)=\{e^{-\IMUN2\pi (\Delta_s+m)/N}|s\in\REALS_{>0},\ m
\in\INTEGERS/N\INTEGERS\}
\]
which coincides with the spectrum of the exponential operator
$e^{-2\pi\IMUN L_0/N}$ in quantum Liouville theory.  Thus, the
discrete version of quantum Liouville theory is in complete agreement
with the continuous one and there is no any modification in the
spectrum of conformal weights.

\end{document}